# Acoustophoresis in Variously Shaped Liquid Droplets

*Gan Yu, Xiaolin Chen, and Jie Xu*\*
*\*Mechanical Engineering, Washington State University, Vancouver, WA, 98686, USA. E-mail: jie.xu@wsu.edu*

The ability to precisely trap, transport and manipulate micrometer-sized objects, including biological cells, DNA-coated microspheres and microorganisms, is very important in life science studies and biomedical applications. In this study, acoustic radiation force in an ultrasonic standing wave field is used for micro-objects manipulation, a technique termed as acoustophoresis. Free surfaces of liquid droplets are used as sound reflectors to confine sound waves inside the droplets. Two techniques were developed for precise control of droplet shapes: edge pinning and hydrophilic/hydrophobic interface pinning. For all tested droplet shapes, including circular, annular and rectangular, our experiments show that polymer micro particles can be manipulated by ultrasound and form into a variety of patterns, for example, concentric rings and radial lines in an annular droplet. The complexity of the pattern increases with increasing frequency, and the observations are in line with simulation results. The acoustic manipulation technique developed here has the potential to be integrated into a more complex on-chip microfluidic circuit. Especially because our method is well compatible with electrowetting technology, which is a powerful tool for manipulating droplets with free surfaces, the combination of the two methods can provide more versatile manipulation abilities and may bring a wealth of novel applications. In the end, we demonstrate for the first time that acoustophoresis can be used for manipulating *Caenorhabditis elegans*.

## Introduction

Particle manipulation, including trapping, transporting, separating and concentrating, is an important task for lab-on-a-chip devices [1-3]. It has been drawing more and more attention in recent years. There are two basic manipulation strategies: one is passively using drag force in a flow [4] or diffusion [5,6] to move particles; the other is actively using other forces, such as electrical forces, magnetic forces and electromagnetic forces (or optical forces), to move particles against the drag force and diffusion. While the first strategy requires sophisticated flow design [7,8], the second strategy often focuses on developing novel ways of controlling various forces, with notable examples of electrophoresis [9-11], dielectrophoresis [12-16], optical tweezers [17-19] and magnetophoresis [20]. It has been found that acoustic radiation force in an ultrasound field can also be used for particle manipulation, a technology termed as acoustophoresis. As an emerging and promising micro manipulation tool, acoustophoresis has many advantages over other methods: it is a non-contact and non-invasive method, and it can work with almost any type of microscale particles regardless of their optical, magnetic or electrical properties. In this paper, we develop novel acoustophoresis techniques for particle manipulation in various shapes of droplets, where the acoustic standing wave field and field-induced radiation force can be precisely controlled. The idea of using droplets for acoustophoresis has been successfully demonstrated by Oberti *et al*, [21]. However, they expressed difficulty in controlling droplet shapes and volumes as well as performing simulations. The shape of the droplet plays a vital role in defining the acoustic standing wave field and simulation is important for system analysis and design. In this paper, we develop two techniques for controlling droplet shapes and demonstrate that a simple 2-D model with impedance boundary conditions can be used to simulate the acoustic standing wave fields. We also demonstrate for the first time that acoustophoresis can be used for manipulating one of the most important model animals in biology - *Caenorhabditis elegans* (*C. elegans*).

### Acoustic Radiation Force

The application of ultrasound, cyclic sound pressure with a frequency greater than the upper audible limit of 20 KHz, in the lab on a chip and microfluidic systems has obtained fast development in recent years. The ultrasound in the form of standing waves will exhibit acoustic radiation force, a non-linear effect that can be used to attract particles to either the nodes or anti-nodes of the standing wave depending on the acoustic contrast factor $\phi$. Inside an acoustic standing wave field, it is the primary acoustic radiation force (PRF) that affects the particles [22]. For a one-dimensional standing planar acoustic wave, the PRF $F_r$ on a spherical particle at the distance $x$ from a pressure node can be calculated by [23]

$$F_r = -\left(\frac{\pi p_0^2 V_p \beta_m}{2\lambda}\right)\phi(\beta,\rho)\sin(2kx) \quad (1)$$

$$\phi(\beta,\rho) = \frac{5\rho_p - 2\rho_m}{2\rho_p + \rho_m} - \frac{\beta_p}{\beta_m} \quad (2)$$

where $\lambda$ is the ultrasonic wavelength, $p_0$ is the pressure oscillating amplitude, $V_p$ is the volume of the particle and $k$ is the wave number. As mentioned above, the acoustic contrast factor $\phi$ defines the force direction and it is a function of the densities and compressibilities of the particle ($\rho_p$, $\beta_p$) and the medium ($\rho_m$, $\beta_m$). According to the equation above, the polymer particles used in this experiment should move towards the pressure nodes [24,25].

### Water Droplets as Acoustic Resonators

Droplet with water/air interface serves as an excellent acoustic resonator in our experiment. According to the theory developed in Ref [23], when an acoustic wave travels from water into air, the first-order velocity potential $\phi_1$ takes the form of Equation (3)

$$\phi_1(x,t) = \begin{cases} \phi_a(x,t) = [A_a e^{ik_a x} + B_a e^{-ik_a x}]e^{-i\omega x}, & \text{in water} \\ \phi_b(x,t) = A_b e^{ik_a x} e^{-i\omega x}, & \text{in air} \end{cases} \quad (3)$$

where $a$ and $b$ indicate water and air, and the coefficients $A_a$, $B_a$ and $A_b$, associated with the incident, reflected and transmitted acoustic waves respectively, can be calculated by Equations (4), (5) and (6)

$$A_a = c_a l \frac{\frac{1}{2}(1+z_{ab})e^{-ik_a l}}{\sin(k_a L) - i z_{ab} \cos(k_a L)} \quad (4)$$

$$B_a = c_a l \frac{\frac{1}{2}(1-z_{ab})e^{ik_a l}}{\sin(k_a L) - i z_{ab} \cos(k_a L)} \quad (5)$$

$$A_b = c_b l \frac{z_{ab} e^{-ik_b l}}{\sin(k_a L) - i z_{ab} \cos(k_a L)} \quad (6)$$

where $c$ is the speed of sound, $l$ is the amplitude of oscillation, $L$ is domain size, and the impedance ratio $z_{ab}$ is defined as

$$z_{ab} \equiv \frac{Z_a}{Z_b} = \frac{\rho_a c_a}{\rho_b c_b} \quad (7)$$

where $\rho$ is the density of material. Table 1 lists these parameters for water and air.

**Table 1** Typical parameter values of air and water at 20℃

| Material | Density $\rho$ [kg m$^{-2}$] | Speed of sound $c$ [m/s] | Impedance $Z$ [kg/(m$^2$ s)] |
| --- | --- | --- | --- |
| Air | $1.2 \times 10^0$ | $3.4 \times 10^2$ | $4.1 \times 10^2$ |
| Water | $1.0 \times 10^3$ | $1.5 \times 10^3$ | $1.5 \times 10^6$ |

From the Equations (4), (5) and (6), we note that if the impedances of two materials are very different, i.e. $z_{ab} \to \infty$, which according to Table 1 is almost the case for the water/air interface, then $|A_a| \approx |B_a| \propto z_{ab} \to \infty$, while $|A_b| \approx C_b l$ which is a finite number. In this situation, most acoustic waves were reflected then confined inside the droplet and only a negligible amount of acoustic waves transmitted through the interface into the air. In this way, the radiation force is optimized by choosing droplets with a water/air interface as acoustic resonators.

## Experimental

### Generation of Acoustic Field and Thermal Control

Ultrasonic standing wave can be generated in one dimension either by the use of two opposing sound sources, or by a single ultrasonic transducer that is facing a sound reflector [22]. In both methods, standing wave is defined by the waves directly produced from the piezoelectric actuators, thus the wave patterns are usually parallel lines. Alternatively, a resonator with a piezoelectric actuator arbitrarily coupled can be used to produce more complex ultrasonic standing waves [24, 26]. In this way, the wave pattern is defined by the specific geometry of the resonator and as a result, the shape of resonator as well as the frequency of the sound will affect the patterns of the acoustic standing wave field, which is studied in-depth in this paper.

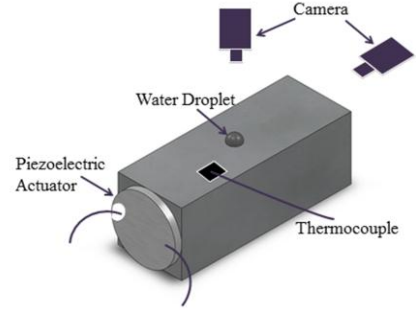

**Fig. 1** The experimental system used in our study. Water droplets are placed on the top surface of the aluminium block for experiments. A piezoelectric actuator attached on the side is used to excite the system with acoustic energy. A high-speed camera is used to observe the experiments either from top or side. A thermocouple is used to monitor the temperature during experiments.

Our experimental system is sketched in Fig. 1. A disk-shaped piezoelectric actuator (HF-28/2MC, Huifeng Piezoelectric Co., LTD) was attached to the side of an aluminium block (27mm×27mm×50mm) using ultrasonic transmission gel (Aquasonic 100, Parker Laboratories, INC). A function generator (DG1022, Rigol) and an amplifier (7602M, Krohn-Hite) were used to send sinusoidal wave signals to the piezoelectric actuator at different frequencies. A thermocouple (Omega K type) was used to monitor the temperature of aluminium block when the piezoelectric actuator is turned on. Water droplets were placed on top the aluminium as acoustic resonators. Top and side views of the droplet can be captured by a high-speed camera (Monochrome Machine Vision Camera, PIxeLINK, PL-B771U).

One obstacle that often limits the application of acoustophoresis in biological studies is the heat generated from the vibration of the piezoelectric actuators: some researchers have to run the actuator at a moderate power level and only for several seconds during experiment [24]. Inspired from Ref [27], we chose a large aluminium block in our experiment and hope it will dissipate the heat fast enough so that no temperature rise can occur during experiments. To prove our intuition, the change in the temperature on the top surface of the aluminium block were monitored and compared to the case where the same piezoelectric actuator was running stand-alone in air. The results are plotted in Fig. 2 for different frequencies.

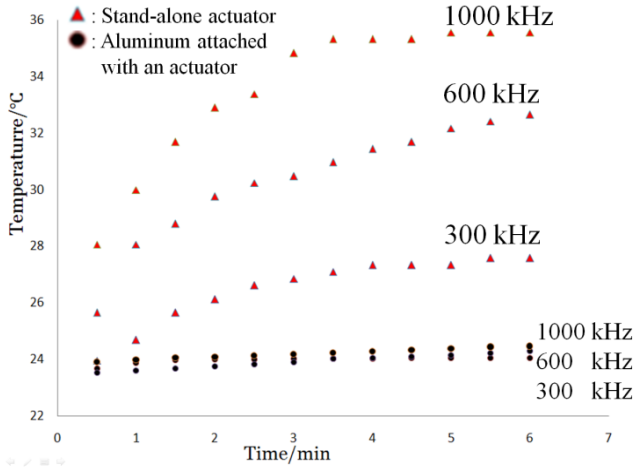

**Fig. 2** Real-time surface temperatures of a stand-alone piezoelectric actuator (red triangle) and an aluminium block attached with the same piezoelectric actuator (black spot). The data points from the aluminium attached with actuator shows negligible change in temperature in all three tested frequencies.

The triangular points represent temperature of the stand-alone piezoelectric actuator and the circular points represent the temperature of the aluminium block attached with the same piezoelectric actuator. Three different frequencies, 300 kHz, 600
5 kHz, and 1000 kHz were tested. From the data, we can see that the temperature of the aluminium almost remains at the room temperature for a long time under different frequencies, while the stand-alone piezoelectric actuator has an obvious increment in temperature, and the higher frequency is, the larger the increment
10 will be.

**Droplet Shape Control**

After the system was built, the next step is to generate droplets
15 with well-defined shapes. In this paper, two techniques were developed to control the shapes of droplets:

1. Hydrophilic/hydrophobic interface pinning

20 An engineered surface with hydrophilic/hydrophobic patterns is a novel way of controlling multiphase interfaces[28-32]. In our experiments, we use the surface of the aluminium as the hydrophilic surface, and use RainX (RainX Original, Sopus Products) coating on aluminium as the hydrophobic surface.
25 Contact angles for water on both surfaces are measured to be $\theta_{aluminum}$=54.2 °± 3.56 ° and $\theta_{RainX}$=109.4 °± 3.91 ° respectively. Different shapes of masks are used during RainX coating on the aluminium surface, so that various hydrophilic/hydrophobic patterns can be created. If a small amount of water was placed
30 onto the surface, it will tend to stay in the hydrophilic areas. If the volume of water increases, by a precision pipette (GeneMate 100, ISCBIOEXPRESS) with 0.04 ml increment, then the drop will grow with a fixed base area and an increasing contact angle $\theta$. Only when the contact angle increases to $\theta_{RainX}$, the droplet can
35 expand over the hydrophilic region. Using this method, circular and triangular droplets were created as demonstrated in Fig 3.

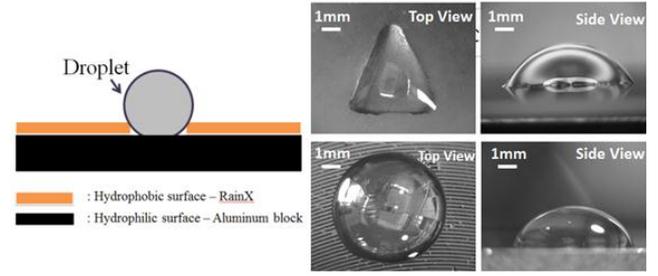

**Fig. 3** Hydrophilic/hydrophobic interface pinning method and designed droplets. Left side is the schematic drawing and right side shows the top and side views of triangular and circular droplets generated by this method.

2. Edge pinning:

40 Alternatively, a physical edge can also be used to pin a contact line, which is used both in nature[33] and in engineered devices[34, 35]. Here we use aluminium protrusions with various base shapes to create corresponding shapes of droplets. As shown in Fig. 4, an
45 annular droplet was created on top of an aluminium o-ring that was attached on top of the aluminium block using ultrasonic transmission gel. A rectangular droplet was also created in this way as seen in Fig. 4.

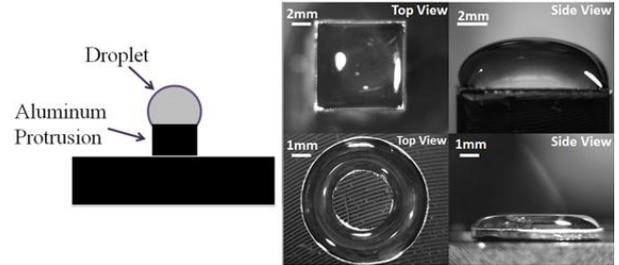

**Fig. 4** Edge pinning method and designed droplets. Left side is the schematic drawing and right side shows the top and side views of rectangular and annular droplets generated by this method.

50 The mechanism of droplet shape control by edge pinning is a little bit more complicated due to the important role that gravity plays, since the size of the aluminium protrusions used in our experiments are usually greater than $\kappa^{-1}$ (= 2.73 mm), the so-called capillary length above which gravity becomes
55 predominate [36]. When the droplet is in equilibrium, the force balance of the droplet gives the following profile function:

$$\gamma + \rho g(ez - \frac{z^2}{z}) = \gamma/\sqrt{1 + \dot{z}^2} \qquad (8)$$

where $\gamma$ (= 73 mN/m) is the surface tension between water and
60 air, $\rho$ is the density of water, $g$ is gravitational constant, $e$ (= 3.86 mm) [36] is the maximum height of the water surface to $x$ axis, $z$ is the profile function of $x$ and $\dot{z}$ is the slope of $z$. A MATLAB code was developed to solve the profile function (8) and the calculated droplet profile is plotted on top of the microscopic pictures taken
65 by the high speed camera in the experiments. In Fig. 5, we can see that the calculated droplet shapes match closely with the actual shapes from the pictures.

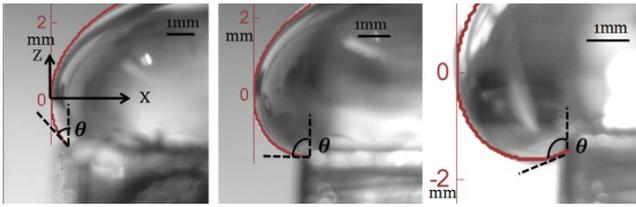

**Fig. 5** Side views of a growing water droplet at an edge of a protrusion. Analytical results of the droplet profile (red lines, calculated from equation 8) are also plotted to compare with experiments. The droplet starts to slip along the vertical wall when the contact angle $\theta$ between the vertical wall and the droplet reaches $\theta_{RainX}$.

The pictures in Fig. 5 from left to right represent the shape of the water droplet changing over time. As the water droplet continues to grow, the contact angle between water and the vertical wall keeps increasing, until the contact angle $\theta$ reaches the $\theta_{RainX}$, at
5 which point the droplet starts to collapse as captured in the last picture in Fig. 5.

While both methods described above can be used to control droplet shapes, each method has its own advantages in future
10 application: the hydrophilic/hydrophobic interface pinning method features planner structure and is then more compatible with the standard microfabrication technology and more suitable for future lab on a chip integration; and the edge pinning method sacrifices the planar structure but gain more ability to pin the
15 contact line, and therefore can be used to fix the droplet base shape over a much wider volume range.

## Acoustophoresis of Particles

After the piezoelectric actuator is turned on, an acoustic field will
20 form inside the droplets. At resonant frequencies, standing wave will start to form and then particles (Copolymer Microsphere Suspension 11μm, Thermo Scientific) inside the droplet can be moved by the acoustic radiation force. The movement of the particles will form in the end different patterns under different
25 frequencies. In the experiments, these patterns were recorded by the camera and simulation results were conducted using ANSYS® software. In the ANSYS simulation, a 2-D acoustic element type Fluid29 was employed for acoustic modal analysis of the pressure field in the simplified droplet chambers. An impedance boundary
30 condition was specified at each air/water interface, with an impedance ratio of water to air given as $Z_{water}/Z_{air} = 1.5 \times 10^6 / 4.1 \times 10^2 = 3658.5$. The following acoustic properties of water were used for the simulation: reference pressure ($1 \times 10^{-6}$ Pa), density (998 kg/m$^3$) and speed of sound (1483m/s). A total of
35 50 acoustic modes (the acoustic standing waves inside the droplet chambers) were extracted over the frequency range of 250-1500 kHz. To ensure result convergence, the mesh densities of the models were increased as frequency goes up to have at least 5 elements per wavelength. Figs. 6-8 report our results.
40
The left columns of Fig. 6 report the experimentally observed particle patterns in a circular droplet, where the white lines are accumulated particles in the acoustic field. These patterns can be categorized into different series. For example, Fig. 6(a) shows a
45 pattern series with only radial lines and Fig. 6(b) is a pattern series with a mix of both radial lines and concentric circles. In both series, the complexity of the pattern increases with increasing frequency, which matches well with the simulation results.
50
The right columns of Fig. 6 report our simulation results, where the resonant pressure contours are plotted. The green areas represent the pressure nodal zones where pressure oscillation is zero, and the blue/red areas represent the anti-nodal zones where
55 pressure oscillation is maximal. The simulated patterns show good consistency with experimental results and prove that the locations where particles gather in the experiments correspond to the nodal zones in the ANSYS results. Similar phenomena can also be observed in annular and rectangular drops as reported in
60 Figs. 7-8.

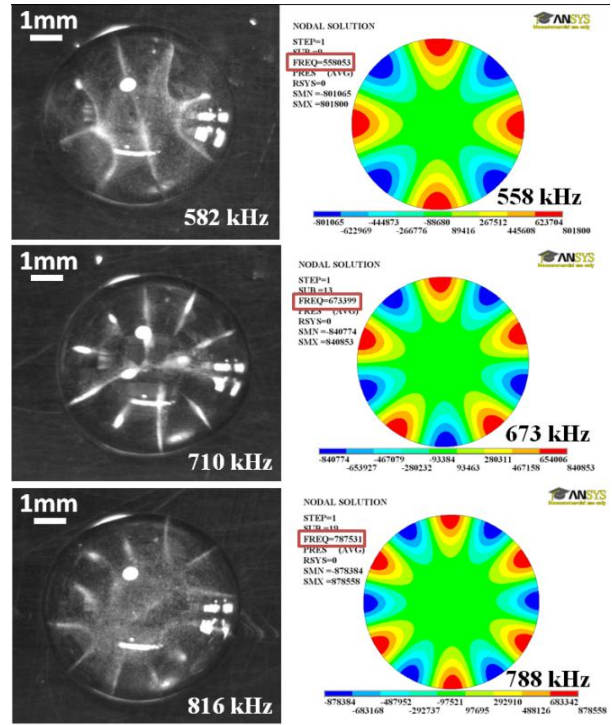

(a)

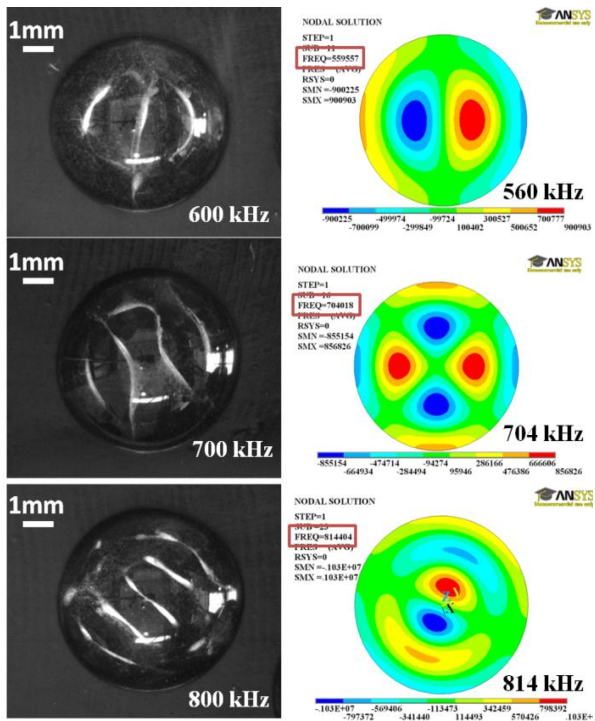

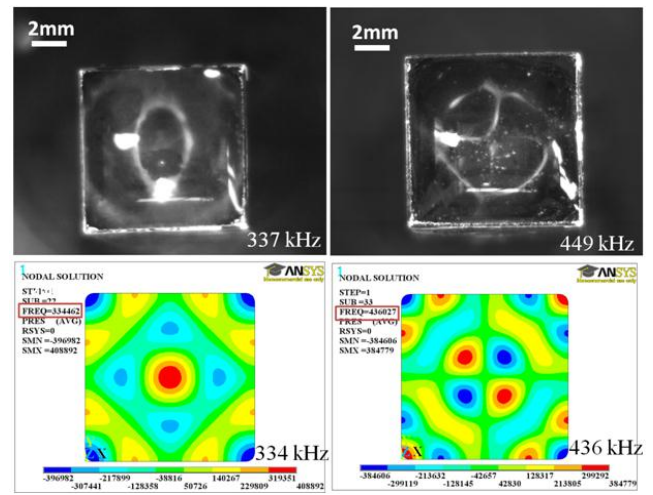

(a)

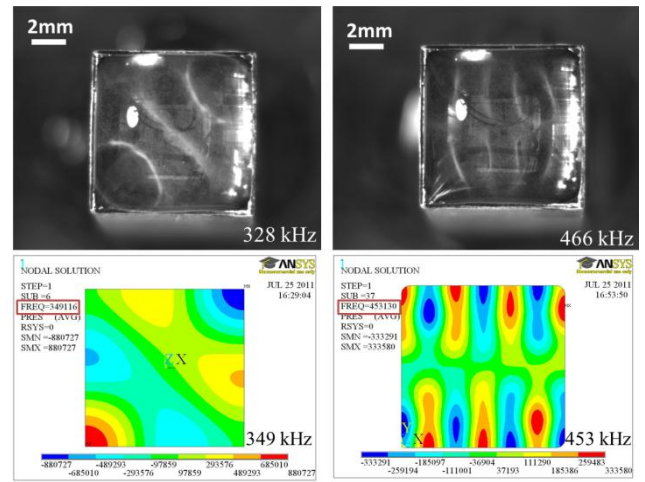

(b)

**Fig. 6** Left columns show the experimental results of circular droplets, where particles form white patterns inside the droplets. Right columns show the corresponding simulation results, the green area in simulation represents the node and the red and blue area represents the anti-node of the resonant acoustic wave. (a): A series of three, five and six radial lines at 582 kHz, 710 kHz and 816 kHz shown on the left and the simulations with same number of radial lines at 558 kHz, 673 kHz and 788 kHz shown on the right. (b): A series of one, two and three lines inside a circle at 600 kHz, 700 kHz and 800 kHz shown on the left and the simulations with same patterns at 560 kHz, 704 kHz and 814kHz shown on the right.

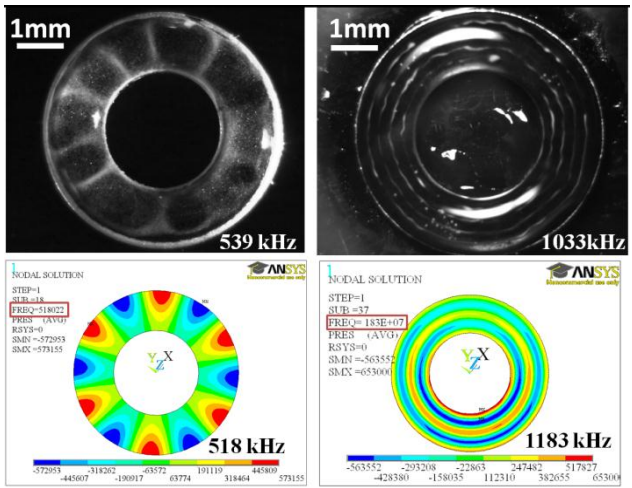

**Fig. 7** Experimental and numerical results with annular droplets. The patterns of fourteen straight lines at 539 kHz and five concentric circles at 1033khz were recorded and simulation results with same patterns at close frequencies were acquired as well.

**Fig. 8** Experimental and numerical results with rectangular droplets. (a): The patterns of one circle at 337 kHz and a circle with a cross at 449 kHz were recorded and simulation results with similar patterns at close frequencies were acquired as well. (b): The patterns of one straight line between two curves at 328 kHz and four vertical lines at 466 kHz were recorded and simulation results with similar patterns at close frequencies were acquired as well.

From the results shown above, we can see that resonant patterns inside a droplet can be affected by both the applied frequency and the base shape of the droplet. In a circular droplet, the symmetry of the pattern seems to be arbitrary, because the base shape has infinite number of axis of symmetry. However, in a rectangular droplet, because the base shape has only four axes of symmetry, the resonant pattern is found to be always symmetrical to one of the axes. Note that, compared to circular droplets, annular droplets tend to show less resonant patterns. This might be due to the missing central area, which limited the possibilities of patterns.

## Acoustophoresis of *C. elegans*

In this section, we explore an application using the acoustophoresis system developed above for manipulating *C.*

*elegans*. *C. elegans* is the first multicellular organism to have its genome completely sequenced [37]. Due to its small size (adult worms are approximately 1mm long), well-mapped neuronal system, transparent body and ease to culture, *C. elegans* is one of the most important model animals in biological and medical research fields. In many studies, these worms need to be selected, sorted or immobilized for observation. Current methods include manually picking up an individual worm and gluing the worm to a surface [38], immobilizing worms in microfluidic channels [39], in which the movement of the worms is typically controlled pneumatically and worms may get hurt sometimes as well. Very recently, it is demonstrated that dielectrophoresis can also be used for worm manipulation [40]. However, the ability to manipulate worms with non-contact forces without electrical fields would be desired for worm study. In our experiments, we confine the *C. elegans* into a 2-D plane for visualization purpose by conveniently modifying the system into a 2-D droplet system as sketched in Fig. 9, where two pieces of glass spacers were used to support another piece of glass and the droplet were then confined inside the gap.

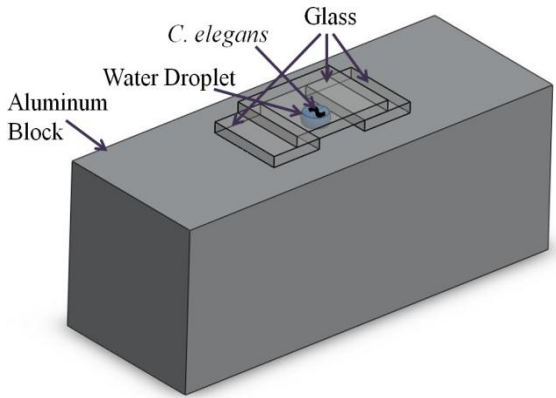

**Fig. 9** 2-D droplet system for manipulating *C. elegans*. Droplet was confined between glass and substrate

During the experiment, a small chunk of NGM (Lab Express) with *C. elegans* (N2 type, requested from Caenorhabditis Genetics Center, University of Minnesota) living inside was first picked up and then immersed into 1 or 2 ml of water, depending on the concentration of worms needed in the experiment. Then the worms will crawl out of the NGM chunk and swim into the water. After half an hour, the water with *C. elegans* inside was collected and then used to create the droplet resonators as shown in Fig. 9. The piezoelectric actuator was applied with a $V_{rms}$ of 122.6 V, the temperature of the living environment of *C. elegans* was 24 °C, and the resonant frequency of the experiment was 1102 kHz. The sequential pictures of how *C. elegans* form into a line is shown in Fig. 10.

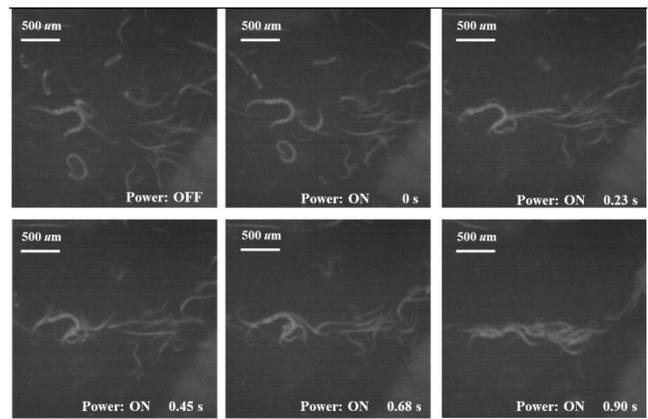

**Fig. 10** When the acoustic field is off, *C. elegans* (white) is distributed evenly inside the droplet. When the acoustic field is applied, *C. elegans* are getting trapped and forming into a line within 1 second.

Fig. 10 shows that before applying the acoustic radiation force, the worms were distributed in the droplet evenly, while after the acoustic radiation force was applied, the worms were gathered by the force into a line instantly (0.9 s) and they could not swim out of that area. As soon as the piezoelectric actuator was turned off, these worms would swim out in random directions and became well dispersed again in the end. During the experiment, the viability of *C. elegans* in the experiment is also tested and shown in Fig. 11.

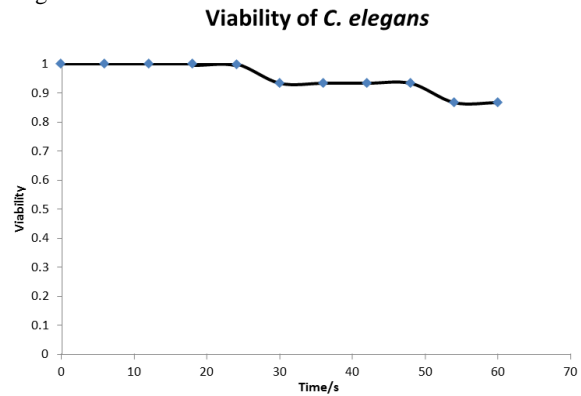

**Fig. 11** The viability (blue dot) of *C. elegans* during 60 s of operation.

Fifteen worms were tested in the experiment and only two died during 60 seconds of operation. The possible reason might be due to the strong shear force exerted on the worm body. However, further study has to be performed to obtain an in-depth understanding.

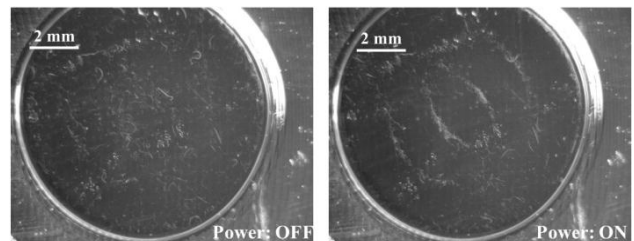

**Fig. 12** High concentration of worms (white) forming into a pattern inside a droplet. (Left) power off and (right) power on.

We have also tested the ability of acoustophoresis to arrange *C. elegans* into patterns. In Fig. 12, a high concentration of *C. elegans* was placed in a droplet. The piezoelectric actuator was applied with a $V_{rms}$ of 106 V, the temperature of the living environment of *C. elegans* is still 24 °C, and the resonant frequency of the experiment is 550 kHz. The left picture in Fig. 11 shows that *C. elegans* were uniformly distributed inside the droplet when the power is off. The right picture in Fig. 11 shows that when the power is turned on, the worms form into a pattern consisting of two circles. The whole process still happens within 1 second.

## Conclusions

In this paper, ultrasonic standing wave was used as a non-contact tool for manipulating objects inside varies shapes of water droplet. Edge pinning and hydrophilic/hydrophobic-interface pinning methods were developed to control the shapes of droplet. Polymer particles were successfully manipulated to form different patterns inside droplets. These observed patterns match the predicted pressure contour patterns from ANSYS simulation. We also report for the first time that *C. elegans* can be manipulated by ultrasonic standing wave using our system. Due to the non-contact and no-electrical-field nature, the technology developed in this paper provides a novel way of manipulating *C. elegans*, as well as potentially any other biological samples. It is worth mentioning that the method presented here is compatible with widely used electrowetting technology[41-43] and switchable-wettability technology[44-49], and the combination of these technologies will provide more versatile manipulation abilities.

## Acknowledgement

We thank Caenorhabditis Genetics Center, for providing us the *C. elegans* strains and Ji Li at Columbia University for helping us in culturing *C. elegans*.